\documentclass[showpacs,aps,pra,superscriptaddress,twocolumn,10pt,amsmath]{revtex4-1}
\usepackage{graphicx}
\usepackage{amsmath}
\usepackage{verbatim}
\usepackage{amssymb}
 
\newcommand{\ket}[1]{\left| #1 \right>}
\newcommand{\bra}[1]{\left< #1 \right|}

\newcommand{\nn}{\nonumber}

\begin{document}

\title{Superadiabatic quantum state transfer in spin chains}
\author{R. R. Agundez}
\email{RRAgundez@gmail.com}
\affiliation{Kavli Institute of Nanoscience, Delft University of Technology, Lorentzweg 1, 2628 CJ Delft, The Netherlands}

\author{C. D. Hill}
\affiliation{Center for Quantum Computation and Communication Technology, School of Physics, The University of Melbourne,Parkville, 3010 VIC, Australia}

\author{L. C. L. Hollenberg}
\affiliation{Center for Quantum Computation and Communication Technology, School of Physics, The University of Melbourne,Parkville, 3010 VIC, Australia}

\author{S. Rogge}
\affiliation{Kavli Institute of Nanoscience, Delft University of Technology, Lorentzweg 1, 2628 CJ Delft, The Netherlands}
\affiliation{Center for Quantum Computation and Communication Technology, School of Physics,The University of New South Wales, Sydney, 2052 NSW, Australia}

\author{M. Blaauboer}
\affiliation{Kavli Institute of Nanoscience, Delft University of Technology, Lorentzweg 1, 2628 CJ Delft, The Netherlands}
\date{\today}

\begin{abstract}
In this article we propose a superadiabatic protocol where quantum state transfer can be achieved with arbitrarily high accuracy and minimal control across long spin chains with an odd number of spins. The quantum state transfer protocol only requires the control of the couplings between the qubits on the edge and the spin chain. We predict fidelities above $0.99$ for an evolution of nanoseconds using typical spin exchange coupling values of $\mu$eV. Furthermore, by building a superadiabatic formalism on top of this protocol, we propose an effective superadiabatic protocol that retains the minimal control over the spin chain and further improves the fidelity.
\end{abstract}

\pacs{03.67.Hk, 75.10.Pq, 03.67.Ac}

\maketitle

\section{Introduction} 
The development of a fully functional quantum computer has been one of the most exciting topics in recent years. An important ingredient is the implementation of a quantum data bus that will transport a desired quantum state with high accuracy. Several types of quantum data buses have been proposed: phonon modes for trapped ions~\cite{Cirac1995,Blatt2008}, cavity photon modes for superconducting qubits and spin qubits~\cite{Imamog1999,Blais2004,Mika2007,Majer2007,Houck2007}, and spin chains for spin qubits~\cite{Bose2003,Friesen2007,Bose2007,Alastair2010,Paganelli2013,Lorenzo2013,Lorenzo2015,Mohi2016}. The latter are a viable candidate for a quantum data bus since the exchange coupling is native to various solid-state systems based on confined electrons. However, the effectiveness of a spin chain as a quantum data bus depends on the availability and controllability of the spin chain couplings. 

In this article we propose a superadiabatic protocol where fast quantum state transfer (QST) can be achieved with high accuracy and minimum control across long spin chains. It has been theoretically shown that by engineering individual couplings in a spin chain QST can be achieved~\cite{Wu2009,Venuti2007,Ashhab2012,Oh2013,Ashhab2015,Zwick2015}. Typically, in these scenarios the qubit-bus couplings are much smaller than the gap between the ground state and the excited state of the spin bus, and QST can be achieved with high precision given initial control 
assumptions~\cite{Oh2011}. However, the state transported is time-dependent and high values of fidelity are only found during small time windows. In comparison, in adiabatic QST the evolution time does not need to be precisely controlled, and once the state is transferred, the system stays in a steady state. Adiabatic protocols are also more viable experimentally because of their robustness against weak variations of the system.

The superadiabatic formalism, in principle, can transform the evolution of any time-dependent Hamiltonian into a purely adiabatic evolution. A recent experiment using this formalism in a system comprising Bose-Einstein condensates in optical lattices gave promising results for future applications in other areas~\cite{Bason2011, Lloyd2012}. Furthermore, recent work on superadiabatic quantum driving has been done in different systems, which illustrates the broad impact of these protocols~\cite{Takahashi2013, Campo2013,Torrontegui2013,Zhang2013,Giannelli2014}. Here we present the first application of superadiabaticity in the context of QST across a spin chain with minimal control.

\begin{figure}
\centering
\includegraphics[width=.8\columnwidth]{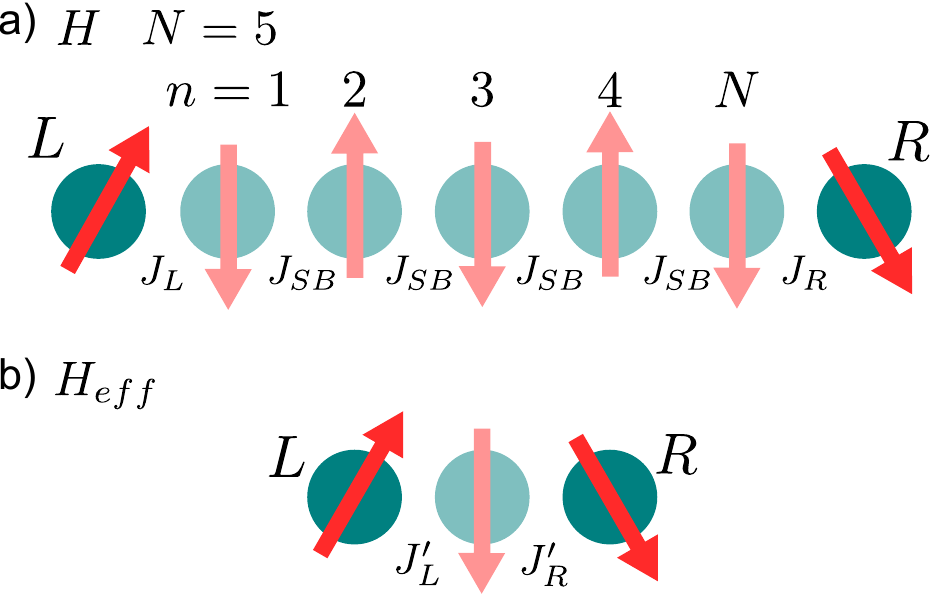}
\caption{a) Schematic of the Heisenberg spin bus with $N=5$ sites [Eq.~(\ref{hamiltonian})]. Our goal is to transfer the state of the sender qubit (L) to the receiver qubit (R). b) Diagram of the system modeled by the effective Hamiltonian for an odd size spin bus [Eq.~(\ref{effectiveHamiltonian})].}\label{fig0} 
\end{figure}

\section{Model} 
The system we consider consists of a Heisenberg spin chain with an odd number $N$ of spins, coupled to an additional qubit at each end (see Fig. \ref{fig0}). We refer to the spin chain as the spin bus, to the left qubit as the sender and to the right qubit as the receiver. We label the sender qubit by $L$ and the receiver qubit by $R$. The Hamiltonian of the system is given by:
\begin{eqnarray}\label{hamiltonian}
H & = & J_L \left( \sigma_L^{x}\otimes\sigma_1^{x} + \sigma_L^{y}\otimes\sigma_1^{y} + \sigma_L^{z}\otimes\sigma_1^{z} \right) + 
\nn\\
& & J_{SB} \sum_{n=1}^{N-1} \left( \sigma_n^{x}\otimes\sigma_{n+1}^{x} + \sigma_n^{y}\otimes\sigma_{n+1}^{y} + \sigma_n^{z}\otimes\sigma_{n+1}^{z} \right) +
\nn \\
& & J_R \left( \sigma_N^{x}\otimes\sigma_R^{x} + \sigma_N^{y}\otimes\sigma_R^{y} + \sigma_N^{z}\otimes\sigma_R^{z} \right),
\end{eqnarray}
where $\sigma_n^x,\sigma_n^y,\sigma_n^z$ are the Pauli matrices for the $n^{th}$ spin in the bus and $J_n$ are the corresponding exchange coupling energies. 

Under the condition that the couplings within the spin bus are much stronger than those to the sender and receiver qubits, $J_{SB}\gg J_R,J_L$, we can treat the first two terms of the Hamiltonian~(\ref{hamiltonian}) as a perturbation. This allows us to project $H$ onto the subspace composed of the sender qubit, the receiver qubit and the spin bus ground states~\cite{Friesen2007,Oh2011}. Adiabatically eliminating higher energy states of the spin chain, we find an effective Hamiltonian  given by, to first order in $J_L$ and $J_R$:
\begin{eqnarray}\label{effectiveHamiltonian}
H_{\rm eft} &=& J_L^{\prime} \left( \sigma_L^{x}\otimes\sigma_{SB}^{x} + \sigma_L^{y}\otimes\sigma_{SB}^{y} + \sigma_L^{z}\otimes\sigma_{SB}^{z} \right) + 
\nn\\
& & J_R^{\prime} \left( \sigma_{SB}^{x}\otimes\sigma_R^{x} + \sigma_{SB}^{y}\otimes\sigma_R^{y} + \sigma_{SB}^{z}\otimes\sigma_R^{z} \right),
\end{eqnarray}
where $\sigma_{SB}^x$, $\sigma_{SB}^y$, $\sigma_{SB}^z$ act on the net spin of the spin bus ground state and the effective couplings are given by the relations:
\begin{eqnarray}\label{effectiveCouplings}
J_L'&=&m_1J_L,\nonumber\\
J_R'&=&m_NJ_R.
\end{eqnarray}
In Eq.~(\ref{effectiveCouplings}) $m_1$ and $m_N$ are the local magnetic moments of the first and last spin in the spin bus respectively. Using $H_{\rm eft}$ thus allows us to describe the $N$-site spin bus as an (effective) spin bus consisting of a single site (under the condition $J_{SB}\gg J_R,J_L$)  \cite{Oh2011}. By adiabatically evolving this Hamiltonian from 
$H_{\rm eft}(t=0)=J_R^{\prime} \left( \sigma_{SB}^{x}\otimes\sigma_R^{x} + \sigma_{SB}^{y}\otimes\sigma_R^{y} + \sigma_{SB}^{z}\otimes\sigma_R^{z} \right)$
to $H_{\rm eft}(t=T)=J_L^{\prime} \left( \sigma_L^{x}\otimes\sigma_{SB}^{x} + \sigma_L^{y}\otimes\sigma_{SB}^{y} + \sigma_L^{z}\otimes\sigma_{SB}^{z} \right)$
the superposition state of the sender qubit is transported to the receiver qubit. QST can then be achieved using a spin bus with an odd number of sites provided that 
$J_{SB}\gg J_R,J_L$ throughout the evolution~\cite{comment0}. 

In preparation for the introduction of the superadiabatic protocol, we first analyse the ordinary adiabatic properties of the three-spin chain. To this end, we parametrize the spin bus-receiver qubit coupling as: $J_R(t)=J_M-J_L(t)$, where $J_L(t)$ evolves from $0$ to $J_M$. This parametrization of $J_R(t)$ gives an effective energy gap $\Delta=2J_M$ between the ground state and the excited state. In particular, we study two different evolutions of $J_L(t)$: a linear evolution given by
\begin{equation}\label{linear}
J_L(t) = \frac{t}{T}J_M,
\end{equation}
and a trigonometric evolution given by
\begin{equation}\label{trigonometric}
	J_L(t)=J_M\sin^2\Big(\frac{\pi t}{2T}\Big).
\end{equation}
This means that the ground state energy is the same at $t=0$ and $t=T$, with the sender qubit and the receiver qubit decoupled respectively. In order for the evolution to be adiabatic, the total evolution time ($T$) is restricted by the condition $T\gg \Delta^{-1}= (2J_M)^{-1}$ \cite{Bacon2009}, which shows that the time it takes to perform the quantum state transfer can be reduced by increasing $J_M$.

To asses the quality of our QST we calculate the fidelity $F$ of the receiver qubit, i.e. at the end of the evolution. $F$ is a measure for how similar the state of the receiver qubit is to the initial superposition state of the sender qubit and is defined as~\cite{Oh2011}:
\begin{equation}\label{fidelity}
F_{\phi}\equiv\bra{\phi}Tr_{L,SB}(\ket{\Psi(t=T)}\bra{\Psi(t=T)})\ket{\phi}.
\end{equation}
Here $\ket{\Psi(t)}$ satisfies the time-dependent Schr\"odinger equation, $H(t)\ket{\Psi(t)}=i\frac{\partial}{\partial t}\ket{\Psi(t)}$, with $H(t)$ the Hamiltonian in Eq.~(\ref{hamiltonian}). $Tr_{L,SB}$ is the trace over the state of the sender qubit and spin bus and $\ket{\phi}=a\ket{\uparrow}+b\ket{\downarrow}$ is the state to be transported. The initial condition is $\ket{\Psi(t=0)}=a\ket{G_0}+b\ket{G_1}$, with $\ket{G_0}$ and $\ket{G_1}$ the degenerate ground states of $H(t=0)$, and we use $a=\exp(i\pi/4)\sin(\pi/4$) and $b=\cos(\pi/4)$~\cite{comment1}.

\begin{figure}
\centering
\includegraphics[width=1.\columnwidth]{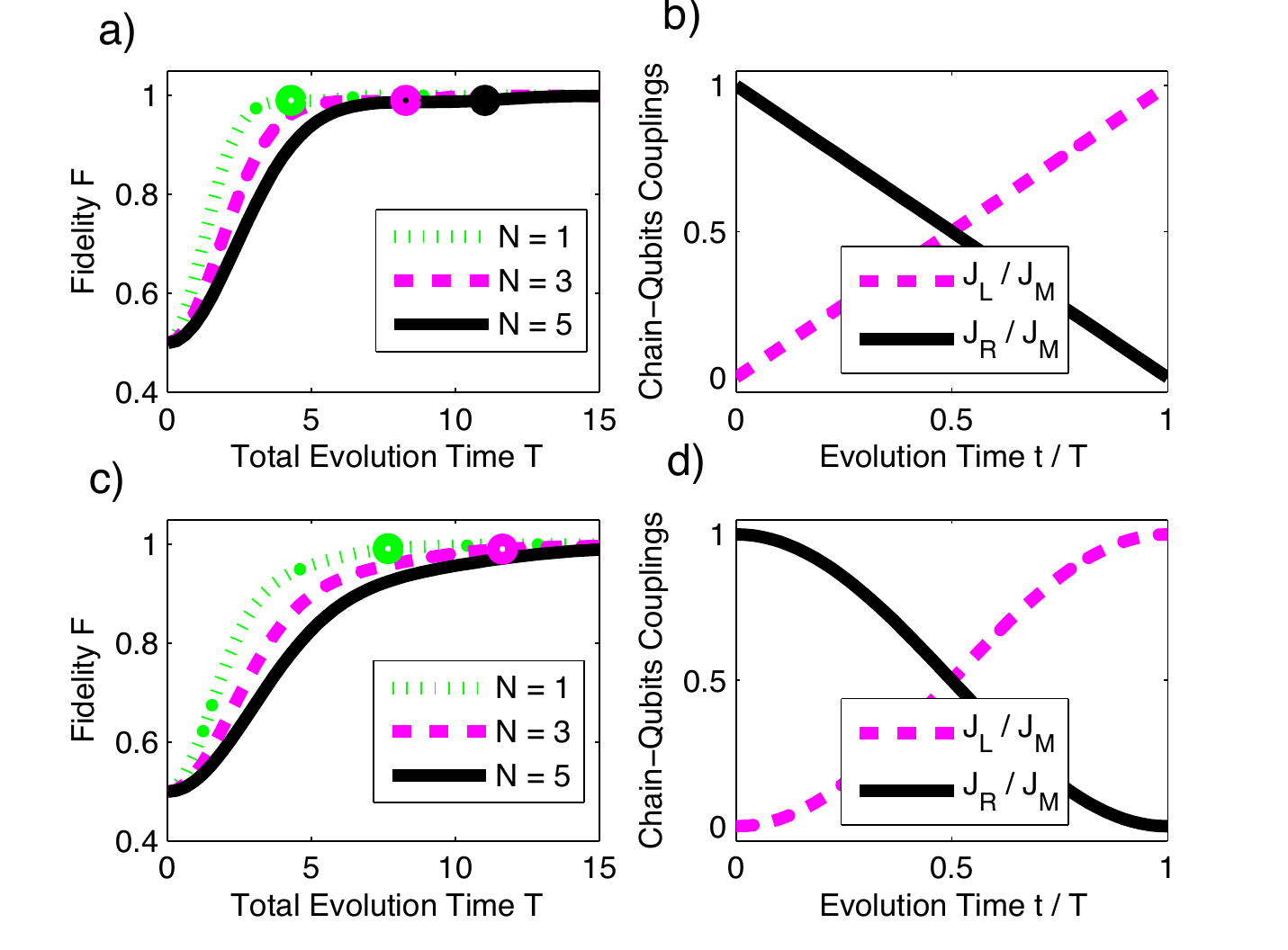}
\caption{Normal adiabatic QST across a spin bus with $N$ sites using the linear and trigonometric time evolution of $J_L(t)$. a) Calculation of the fidelity $F$ [Eq.~(\ref{fidelity})] versus the total evolution time $T$ (in units of $\Delta^{-1}$) for the linear evolution of $J_L(t)$ [Eq.~(\ref{linear})]. b) The linear evolution of $J_L(t)$ induces a linear evolution of $J_{R}$. c) Fidelity $F$ versus the total evolution time $T$, for the trigonometric evolution of $J_L(t)$ [Eq.~(\ref{trigonometric})]. d) Time evolution of the couplings used in c). In these calculations $J_M = 1$ and $J_{SB} = 10$. The points from where onwards fidelity reaches high values ($F > 0.99$) are denoted by circles.}\label{fig2} 
\end{figure} 

The results of the (normal adiabatic) QST using both evolutions are shown in Fig.~\ref{fig2}. In Fig.~\ref{fig2}(a) we have plotted the fidelity at the end of the linear protocol for different spin bus sizes. We observe that QST is achieved with fidelities $F>0.99$ for each spin bus (indicated by circles in Fig.~\ref{fig2}(a)). Larger spin buses require a longer evolution time to achieve QST since the gap decreases as $\Delta\propto \frac{1}{N}$~\cite{Oh2011}. A smaller gap requires a slower evolution to achieve adiabaticity ($T\gg 1/\Delta$).

The linear evolution is smoother and has a smaller derivative at $t = T/2$ where the minimum gap ($\Delta$) between the ground state and the excited state is found [Figs. \ref{fig2}(a)-\ref{fig2}(c)]. The circles in Fig. \ref{fig2}(c) show that fidelities for trigonometric evolution are lower than for the linear evolution in Fig.~\ref{fig2}(a). Moreover, still high fidelities can be achieved if the system is let to evolve slower.

\begin{figure}[h]
\centering
\includegraphics[width=1.\columnwidth,height=6.5cm]{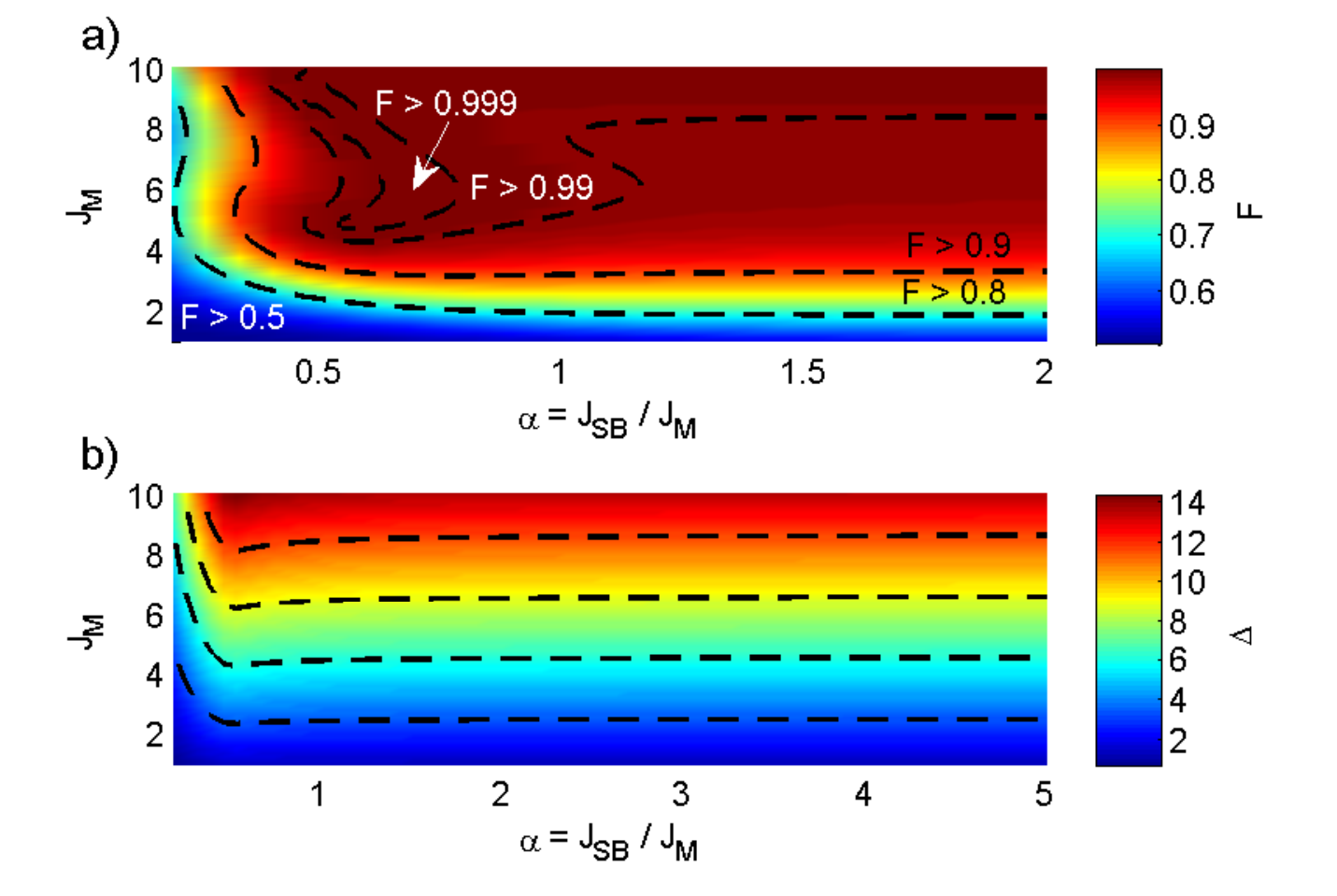}
\caption{a) Contour plot of fidelity $F$ [Eq.~(\ref{fidelity})] versus $J_M$ and $\alpha=\frac{J_{SB}}{J_M}$. The region $\alpha>1$ can be described by the effective Hamiltonian [Eq.~(\ref{effectiveHamiltonian})]. A high fidelity region appears for $\alpha<1$ (see text for discussion). b) Contour plot of the energy gap between the ground state and the first excited state for the same parameters as in a). A spin bus with $N=3$ and linear time evolution with $T=1\, \Delta^{-1}$ was used in both figures a) and b).}\label{fig3} 
\end{figure}

As larger $J_M$ will increase $\Delta$, the evolution time necessary to achieve high fidelities should decrease. On the other hand, increasing $J_M$ will increase the maximum value of the couplings $J_L(t)$ and $J_R(t)$ 
[Eqs.~(\ref{linear})-(\ref{trigonometric})]. Therefore, to maintain the condition $J_{SB}\gg J_R,J_L$, we also need to increase $J_{SB}$. Defining $\alpha\equiv\frac{J_{SB}}{J_M}$ we now study how $F$ depends on the (relative) increase of $J_{SB}$ and $J_M$. For large $\alpha$ the system should approach the effective Hamiltonian behavior.  

Fig.~\ref{fig3}(a) shows a contour plot of $F$ versus $J_M$ and $\alpha$ for the linear evolution with $T=1\, \Delta^{-1}$ in a $N=3$ spin bus. In the region $\alpha>1$ the value of the fidelity no longer depends on $J_{SB}$ as predicted by the effective Hamiltonian Eq.~(\ref{effectiveHamiltonian}). Moreover we find that a higher value of $J_M$ increases the fidelity of the QST. In particular, Fig.~\ref{fig3}(a) shows that the fidelity increases from $F\approx 0.53$ to $F\approx 0.999$ when $J_M$ increases from $1$ to $10$, given $\alpha>1.4$. The increase in fidelity is a consequence of a larger gap between ground state and excited state as shown in Fig. \ref{fig3}(b) ($\Delta = 2J_M$).

In the regime not modeled by the effective Hamiltonian ($\alpha<1$ in Fig.~\ref{fig3}(a)) we find an interesting region where high fidelities can be achieved ($F>0.999$). This high fidelity region is not a consequence of an increment in the gap between ground and excited state [Fig.~\ref{fig3}(b)]. We suspect that these high fidelities originate from the fact that the energy of the first excited state rapidly varies with time for low values of alpha and becomes time-independent when the fidelity is at its maximum value. The increase of fidelity with $J_M$ and the appearance of a high fidelity region for $\alpha<1$ is also observed in longer spin buses ($N=5$ and $N=7$) and for the trigonometric time evolution. 

\section{Effective superadiabatic model} 
Given a time dependent Hamiltonian, $H_{\rm eft}(t)$, it is possible to construct a Hamiltonian $H_{\rm eft,s}(t)$ such that it cancels the non-adiabatic evolution of $H_{\rm eft}(t)$. Then the evolution of the so-called superadiabatic Hamiltonian $H_{\rm eft,S}(t)=H_{\rm eft}(t)+H_{\rm eft,s}(t)$, is ensured to be completely adiabatic. The Hamiltonian $H_{\rm eft,s}(t)$ is calculated as~\cite{Bason2011}:
\begin{equation}\label{counterDiabatic}
	H_{\rm eft,s}(t)=\frac{i}{2}\sum_{\psi}{\Big[\frac{\partial}{\partial_t}(\ket{\psi(t)})\bra{\psi(t)}-\ket{\psi(t)}\frac{\partial}{\partial_t}(\bra{\psi(t)})\Big]},
\end{equation}
where $\ket{\psi(t)}$ are the eigenstates of the Hamiltonian $H_{\rm eft}(t)$.

We could treat the complete spin bus architecture $H$ [Eq.~(\ref{hamiltonian})] with the superadiabatic formalism, but this has several drawbacks: the couplings within the spin bus will become time-dependent, the derivation of the superadiabatic Hamiltonian will be $N$-specific and imaginary matrix elements will be introduced [Eq.~(\ref{counterDiabatic})]. Furthermore, these superadiabatic terms consist of many-body interactions which increases the complexity of converting them into experimental knobs as $N$ increases. Instead, we work with the effective Hamiltonian and then map the couplings onto the complete Hamiltonian keeping $J_{SB}\gg J_L,J_R$, which, as we will show, eliminates every drawback and reduces the number of superadiabatic many body interaction terms.

We find that $H_{\rm eft,S}$ contains time-dependent interactions, coupling the states $\ket{\uparrow\uparrow\downarrow}$ and $\ket{\downarrow\uparrow\uparrow}$, and $\ket{\uparrow\downarrow\downarrow}$ and $\ket{\downarrow\downarrow\uparrow}$. These correspond to an interaction proportional to $\sigma_L^{x}\otimes\sigma_R^{x} + \sigma_L^{y}\otimes\sigma_R^{y} + \sigma_L^{z}\otimes\sigma_R^{z}$, which appears in the second-order term of the effective Hamiltonian and is $J_{SB}$-dependent  \cite{Friesen2007,Oh2011}. Since we want $J_{SB}$ to be time-independent in order to allow for minimal control we neglect these matrix elements. As a result, the evolution will not be completely adiabatic. Nevertheless we expect to improve the QST fidelity due to the other superadiabatic matrix elements retained that improve adiabaticity. We call this Hamiltonian "effective superadiabatic" and denote it by $H_{\rm eft,ES}$. Since the matrix elements in $H_{\rm eft,ES}$ originating from $H_{\rm eft,s}$ [Eq.~(\ref{counterDiabatic})] are imaginary, we apply a unitary transformation $U$ such that $H'_{\rm eft,ES} = UH_{\rm eft,ES}U^\dagger$ has only real matrix elements that we can then rewrite as spin exchange interactions. In particular, we choose the unitary transformation
\begin{equation}
	U(t) = \exp[-\frac{i}{2}\theta_L(t)\sigma^z_L\sigma^z_{SB}-\frac{i}{2}\theta_R(t)\sigma^z_{SB}\sigma^z_R],
\end{equation}
which corresponds to a rotated frame with correlated phases between the central spin and the sender and receiver qubits. $U$ is determined so that $H'_{\rm eft,ES}$ has the same form as $H_{\rm eft}$ [Eq. (\ref{effectiveHamiltonian})]. This imposes the following conditions on $\theta_L(t)$ and $\theta_R(t)$:
\begin{eqnarray}
	\tan{\theta_L} &=& \frac{J_M\frac{\partial J_L}{\partial t}}{4(2J_M-2J_L)(3J_L^2+J_M^2-3J_MJ_L)}, \\
	\tan{\theta_R} &=& \frac{-J_M\frac{\partial J_L}{\partial t}}{8J_L(3J_L^2+J_M^2-3J_MJ_L)}.
\end{eqnarray}
The new transformed state $\ket{\psi'(t)}=U(t)\ket{\psi(t)}$ obeys the Schr\"odinger equation with the Hamiltonian:
\begin{eqnarray}\label{Hamiltonian_ps}
\tilde{H}_{\rm eft,ES}(t) & = & \tilde{J}_L \left( \sigma_L^{x}\otimes\sigma_{SB}^{x} + \sigma_L^{y}\otimes\sigma_{SB}^{y} + \sigma_L^{z}\otimes\sigma_{SB}^{z} \right) + 
\nn\\
& & \tilde{J}_R \left( \sigma_{SB}^{x}\otimes\sigma_R^{x} + \sigma_{SB}^{y}\otimes\sigma_R^{y} + \sigma_{SB}^{z}\otimes\sigma_R^{z} \right),
\end{eqnarray}
where 
\begin{eqnarray}\label{couplings3}
\tilde{J}_L(t)&=&(J_L\sec\theta_R, J_L\sec\theta_R, J_L+\frac{1}{2}\frac{\partial\theta_L}{\partial t})
\nn \\
\tilde{J}_R(t)&=&(J_R\sec\theta_L, J_R\sec\theta_L, J_R+\frac{1}{2}\frac{\partial\theta_R}{\partial t}).
\nn \\
\end{eqnarray}
The Hamiltonian to apply in the lab frame is the original Hamiltonian $H$ from Eq. (\ref{hamiltonian}), using the new edge couplings from Eq. (\ref{couplings3}) and keeping $J_{SB}>J_M$.  The initial value of $\tilde{J}_L$ for the linear and trigonometric evolution are equal to 
\begin{eqnarray}
\tilde{J}_{L,lin}(0)&=&(\frac{1}{8T},\frac{1}{8T},\frac{4J_M(3J_MT+1)}{1+(8J_MT)^2}) \label{coupling_linear}\\
\tilde{J}_{L,trig}(0)&=&(0,0,\frac{\pi^2}{32J_MT^2}) \label{coupling_trig}.
\end{eqnarray}
For both evolutions at least one of the components is unbound for decreasing $T$; we then do not expect to find high fidelities for short evolution times.
The $x$- and $y$-components of the couplings at $t=0$ and $t=T$ are proportional to $\frac{\partial J}{\partial t}$, while the $z$-component has a contribution from $\frac{\partial^2 J}{\partial t^2}$.
Hence having both derivatives equal to zero at these times is desirable for achieving high fidelities and fast QST. This is why the linear evolution produces couplings with large $x$- and $y$-components and a small $z$-component, while the trigonometric evolution produces $x$- and $y$-components equal to zero and a large $z$-component for small $T$. From Eqns.~(\ref{coupling_linear}) and~(\ref{coupling_trig}) we also notice that $T$ not only affects the adiabaticity requirement ($T \gg \Delta^{-1}$) but also the QST protocol itself.

\begin{figure}
\centering
\includegraphics[width=1.\columnwidth,height=9.cm]{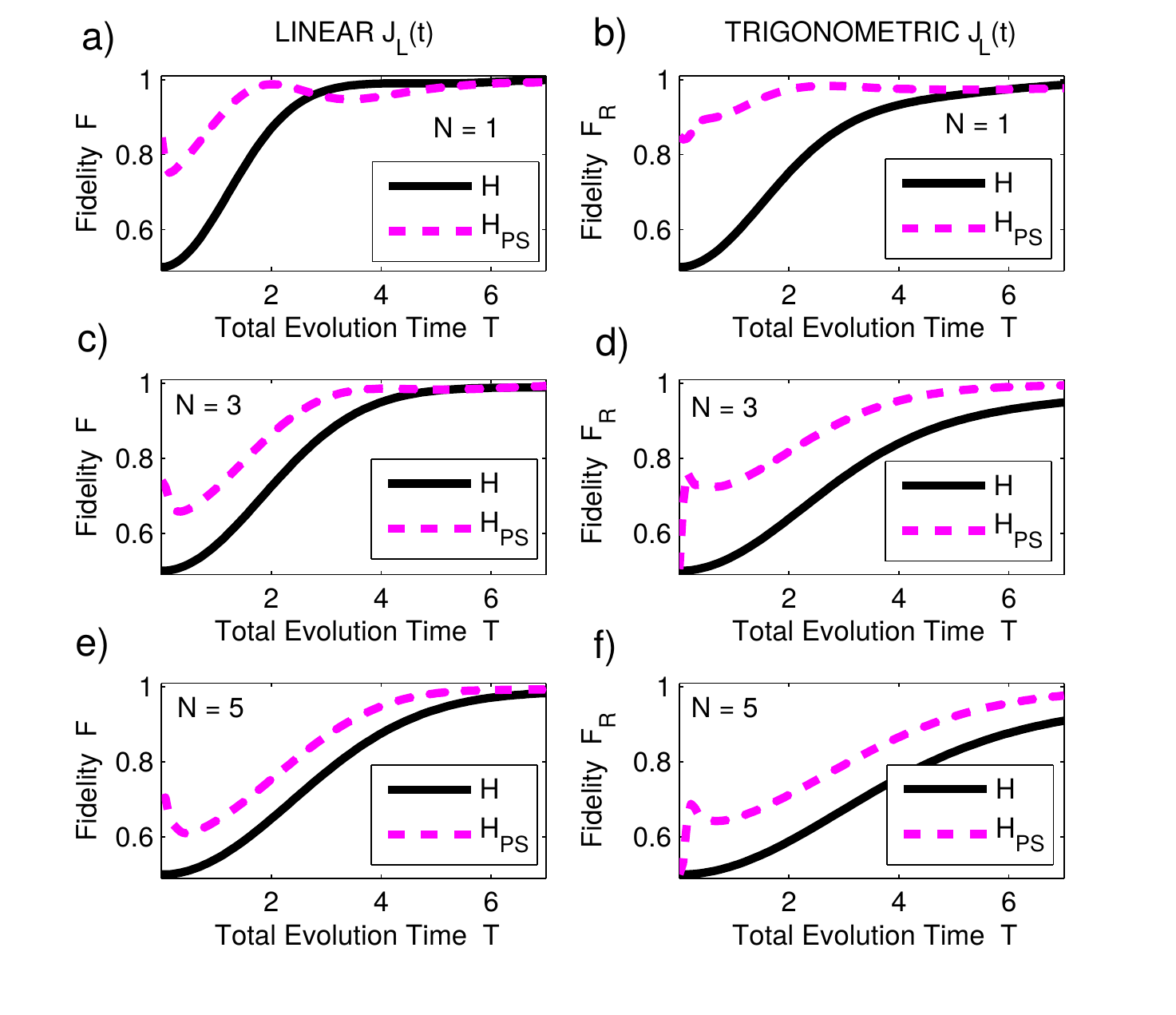}
\caption{Fidelity $F$ for the linear and trigonometric evolution of $J_L(t)$ as a function of $T$ (in units of $\Delta^{-1}$) for spin bus sizes $N=1,3,5$. Values of $J_M=1$ and $J_{SB}=10$ were used. The black-solid lines correspond to a normal adiabatic QST using the Hamiltonian~(\ref{hamiltonian}), the magenta-dashed lines are for the effective superadiabatic Hamiltonian~(\ref{Hamiltonian_ps}). In every case higher fidelities can be seen using the effective superadiabatic quantum state transfer.}\label{fig7} 
\end{figure}

The results from QST calculations for spin buses with $N=1,3,5$ are shown in Fig.~\ref{fig7}. We show a comparison between the quantum state transfer using the original Hamiltonian $H$ 
[Eq.~(\ref{hamiltonian})] and the effective superadiabatic Hamiltonian $H_{ES}$ [Eq.~(\ref{hamiltonian}) with $J_L=\tilde{J}_L$ and $J_R=\tilde{J}_R$]. We observe that higher fidelities are achieved using the effective superadiabatic protocol for most QST durations $T$. Using the trigonometric evolution of $J_L(t)$ the fidelity is lower for $T\rightarrow 0$, because the coupling scales as $\propto \frac{1}{T^2}$ [Eq.~(\ref{coupling_trig})] in contrast with the linear evolution that scales as $\propto \frac{1}{T}$ [Eq.~(\ref{coupling_linear})]. We see that the effective superadiabatic trigonometric evolution achieves an improvement for any evolution time $T$ in long spin buses. Comparing F for different spin bus sizes suggests that the improvement increases with spin bus size.

\begin{figure}
\centering
\includegraphics[width=.9\columnwidth]{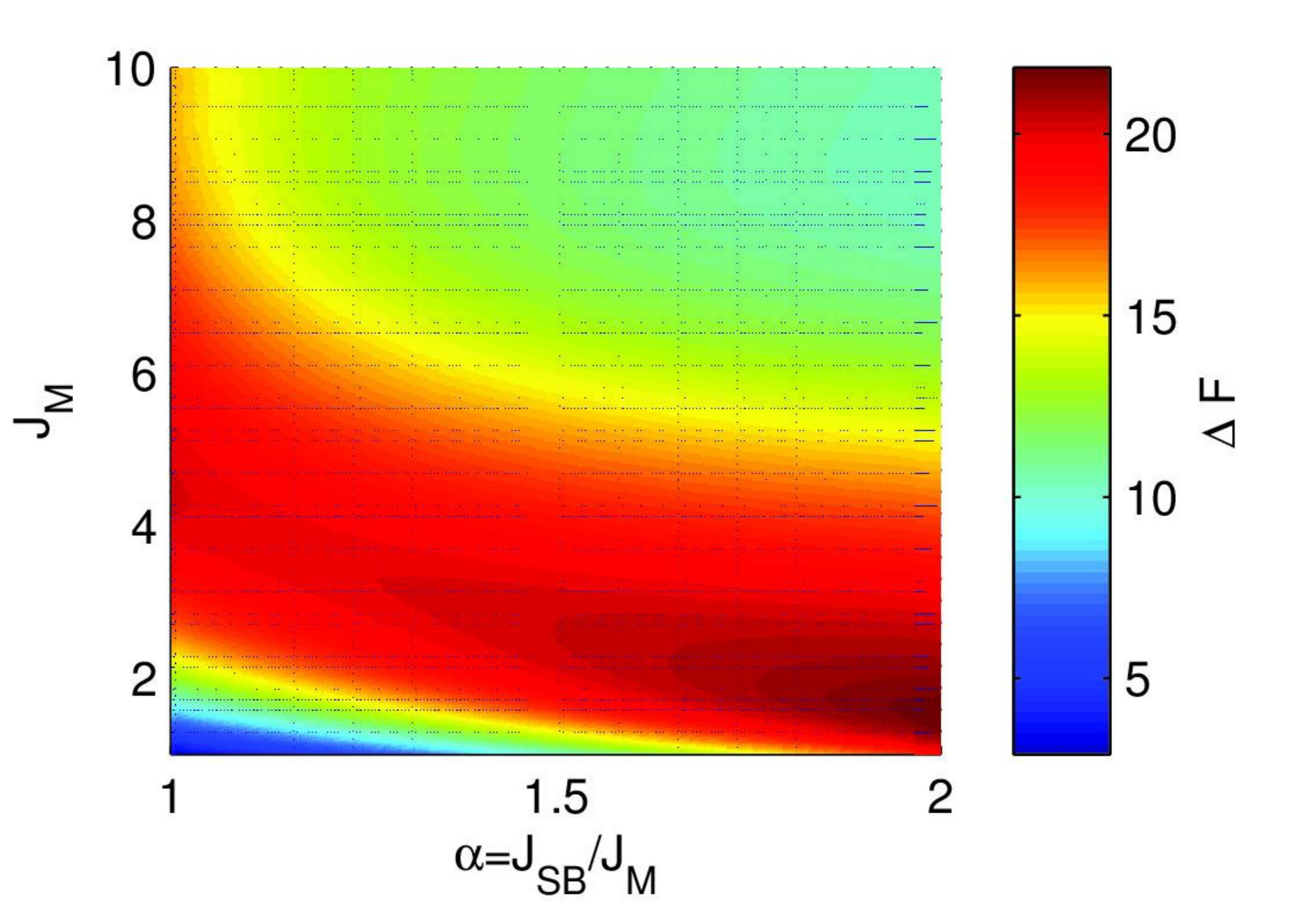}
\caption{Surf plot of the improvement of fidelity due to the effective superadiabatic Hamiltonian for the trigonometric evolution of $J_L(t)$ and the region $\alpha>1$ described by the effective Hamiltonian. In this calculation $N=5$ and $T=1\, \Delta^{-1}$.}\label{fig8} 
\end{figure}

Both protocols did not improve the fidelity anywhere in the region $\alpha=J_{SB}/J_M<1$ (as expected, since our derivations are only applicable for $\alpha>1$). To quantify the improvement of the effective superadiabatic protocol we introduce the following quantity
\begin{equation}
\Delta F= \frac{F_{ES}-F}{1-F}\times 100\%.
\label{eq:DeltaF}
\end{equation}
$\Delta F$ measures the fidelity improvement by the effective superadiabatic protocol using $H_{ES}$ [Eq. (\ref{Hamiltonian_ps})] as a percentage of the difference between the fidelity of 
perfect QST ($F=1$) and the fidelity of the normal adiabatic QST protocol using $H$ [Eq.~(\ref{hamiltonian})]. Here $F_{ES}$ denotes the fidelity that is obtained by using the effective superadiabatic protocol. A value of $\Delta F=100\%$ corresponds to the case where the effective superadiabatic protocol improves the fidelity up to unity. Fig.~\ref{fig8} shows $\Delta F$ for a spin bus consisting of 5 sites and trigonometric evolution of $J_L(t)$. We notice that the improvement can reach up to 25\% (unfortunately this occurs in a region where fidelities are not very high [Fig.~\ref{fig3}(a)]). In the high fidelity region ($J_M>7$) $\Delta F \sim$ 10-15\%. Although this may not seem such a big difference, it is a significant improvement when keeping in mind that the effective superadiabatic protocol is achieving an improvement of a QST with already very high fidelities ($F>0.99$). 
Taking into account that $\hbar\approx 1$ $\mu$eV$\cdot$ns, for values of $J_M=1$ $\mu$eV (well within reach for eg. coupled donors in silicon and electron spins in quantum dots~\cite{Koiller2004,Mohi2016}) we predict a QST duration of a few nanoseconds.

\section{Conclusion} 
In this article we develop a high fidelity superadiabatic quantum state transfer protocol for a spin bus architecture with an odd number of spins. 
The proposed protocol can be implemented with minimal control of the nearest-neighbour exchange couplings long the chain: only the
coupling $J_L(t)$ between the sender qubit and the first spin at one end of the chain, and the coupling $J_R(t)$ between the 
last spin and the receiver qubit at the other end need to be controlled. We apply this protocol via an effective transformation and 
consider two types of evolution of $J_L(t)$ and $J_R(t)$, linear and trigonometric, and find that for both the superadiabatic protocol leads to 
a significant relative increase, quantified by $\Delta F$ in Eq.~(\ref{eq:DeltaF}), in the fidelity of QST of up to 25\% compared to normal adiabatic evolution.
Comparing fidelities for different spin chain lengths suggests that this improvement increases with spin bus size.
\vspace*{0.5cm}
\section{acknowledgments}
This work is part of the research program of the Foundation for Fundamental Research on Matter (FOM), which is part of the Netherlands Organization for Scientific Research (NWO) and was financially supported by the ARC Centre of Excellence for Quantum Computation and Communication Technology (CE110001027) and the Future Fellowship (FT100100589).


\begin{thebibliography}{99}
\bibitem{comment1}
These values of $a$ and $b$ are arbitrary choices; the same results are observed for any other values.
\end{thebibliography}
\end{document}